%% file: paper_MCS_JSS_v3_arxiv.tex
\documentclass[nojss]{jss}

\usepackage{amsfonts}
\usepackage{amsmath}
\usepackage{amssymb}
\usepackage{cleveref}

\usepackage{multicol}
\usepackage{multirow}
\usepackage{booktabs}
\usepackage{rotating}
\usepackage{graphicx}
\usepackage{bbm}
\usepackage[round]{natbib}

\newcommand\bbone{\ensuremath{\mathbbm{1}}}
\input{mydef.tex}

\ifpdf
    \graphicspath{{Figs/Raster/}{Figs/PDF/}{Figs/}}
\else
    \graphicspath{{Figs/Vector/}{Figs/}}
\fi

\author{Mauro Bernardi\\Sapienza University of Rome \And
        Leopoldo Catania\\University of Rome Tor Vergata}
\title{The Model Confidence Set package for \proglang{R}}

\Plainauthor{Mauro Bernardi, First Author} %% comma-separated
\Plaintitle{The Model Confidence Set package for R} %% without formatting
\Shorttitle{The MCS package for R} %% a short title (if necessary)

\Abstract{This paper presents the \proglang{R} package \code{MCS} which implements the Model Confidence Set (MCS) procedure recently developed by \cite{hansen_etal.2011}. The Hansen's procedure consists on a sequence of tests which permits to construct a set of \qmo superior\qmcsp models, where the null hypothesis of Equal Predictive Ability (EPA) is not rejected at a certain confidence level. The EPA statistic tests is calculated for an arbitrary loss function, meaning that we could test models on various aspects, for example punctual forecasts. The relevance of the package is shown using an example which aims at illustrating in details the use of the functions provided by the package. The example compares the ability of different models belonging to the ARCH family to predict large financial losses. We also discuss the implementation of the ARCH--type models and their maximum likelihood estimation using the popular \proglang{R} package \code{rugarch} developed by \cite{ghalanos.2014}.}
\Plainkeywords{Hypothesis testing, Model Confidence Set, Value--at--Risk, VaR combination, ARCH--Models, R--CRAN} %% without formatting
\Address{
  Mauro Bernardi\\
  Department of Methods and Models for Economics Territory and Finance\\
  Faculty of Economics\\
  Sapienza University of Rome\\
  Via del Castro Laurenziano, 9\\
  00161 Rome, Italy\\
  E-mail: \email{mauro.bernardi@uniroma1.it}\\
  Leopoldo Catania\\
  Department of Economics and Finance\\
  Faculty of Economics\\
  University of Rome Tor Vergata\\
  Via Columbia, 2\\
  00133 Rome, Italy\\
  E-mail: \email{leopoldo.catania@uniroma2.it}\\
}

\begin{document}

%% include your article here, just as usual
%% Note that you should use the \pkg{}, \proglang{} and \code{} commands.

%:::::::::::::::::::::::::::::::::::::::::::::::::::::::::::::::::::::
% SECTION: INTRODUCTION
%:::::::::::::::::::::::::::::::::::::::::::::::::::::::::::::::::::::
\section{Introduction}
\label{sec:intro}
%:::::::::::::::::::::::::::::::::::::::::::::::::::::::::::::::::::::
%
During last decades hundred of models have been developed, estimated and validated both from an empirical and theoretical point of view with the consequence that usually several competing models are available to the econometricians to address the same empirical problem. Just to confine our considerations within a given family, and without claiming to be complete, the Autoregressive Conditional Heteroskedastic (ARCH) models of \cite{engle.1982} and \cite{bollerslev.1986} have seen an exponentially increasing number of different specifications in the last few decades. Despite their popularity, they do not exhaust the set of models introduced for dynamic conditional volatility modelling which includes also the stochastic volatility models initially proposed by \cite{taylor.1994} and extensively studied by \cite{harvey_shephard.1996} and \cite{gallant_etal.1997} within the context of non--linear state space models. The family of dynamic conditional volatility models has been recently enlarged by the Dynamic Conditional Score models of \cite{harvey.2013} and \cite{creal_etal.2013} also known as Generalised Autoregressive Score models. The availability of such an enormous number of models raises the question of providing a statistical method or procedure that delivers the \qmo best\qmcsp models with respect to a given criterium. Furthermore, a model selection issue appears when the usual comparing procedures does not deliver an unique result. This could happen for example, when models are compared in terms of their predictive ability so that models that produce better forecasts are preferred. Unfortunately, sometimes it is not trivial to asses which model clearly outperforms each other. This problem is relevant from an empirical point of view especially when the set of competing alternatives is large. As observed by \cite{hansen_lunde.2005} and \cite{hansen_etal.2011}, since different competing model are usually built to answer a specific econometric question, we not expect that a single model dominates all competitors either because they are statistically equivalent or because there is not enough information coming from the data to univocally discriminate the models. Recently, a lot of effort has been devoted to develop new testing procedures being able to deliver %an entire set of models that contains
the \qmo best fitting\qmcsp models, see e.g. the Reality Check (RC) of \citet{white.2000}, the Stepwise Multiple Testing procedure of \cite{romano_wolf.2005} and the Superior Predictive Ability (SPA) test of \citet{hansen_lunde.2005} and the Conditional Predictive Ability (CPA) test of \cite{giacomini_white.2006}.\newline
%%%
Among those multiple--testing procedures, the Model Confidence Set procedure (MCS) of \cite{hansen_etal.2003}, \cite{hansen_reinhard.2005} and \cite{hansen_etal.2011} consists of a sequence of statistic tests which permits to construct a set of \qmo superior\qmcsp models, the \qmo Superior Set Models\qmcsp (SSM), where the null hypothesis of equal predictive ability (EPA) is not rejected at certain confidence level $\alpha$. The EPA statistic tests is evaluated for an arbitrary loss function, which essentially means that it is possible to test models on various aspects depending on the chosen loss function. The possibility of user supplied loss functions provides enough flexibility to the procedure that can be used to test competing models with respect to several different aspects. The Model Confidence Set procedure starts from an initial set of $m$ competing models $\mathrm{M}_{0}$ and results in a (hopefully) smaller set of superior models (i.e. SSM) denoted by $\mathrm{\widehat{M}}_{1-\alpha}^{*}$; of course the best scenario is when the final set consists of a single model. At each step of the iterative procedure, the EPA hypothesis is tested, and if the null hypothesis is accepted, then the procedure stops and the SSM is created, otherwise, the EPA should be tested again after the elimination of worst model.\newline
%%%
The models belonging to the superior set delivered by the Hansen's procedure can then be used for different purposes. For example, they can be used to forecast future volatility level, to predict the future level of observations, conditional to past information, or to deliver future Value--at--Risk levels as argued by \cite{bernardi_etal.2014}. Alternatively, those models can be combined together to obtain better forecast measures. Since the original work of \cite{bates_granger.1969}, a lot of papers have argued that combining predictions from alternative models often improves upon forecasts based on a single \qmo best\qmcsp model. In an environment where observations are subject to structural breaks and models are subject to different levels of misspecification, a strategy that pools information coming from different models typically performs better than methods that try to select the best forecasting model. In the empirical part of the paper, where several alternative GARCH specifications are compared in terms of their ability to predict future large losses, we also consider the Dynamic Model Averaging technique proposed by \cite{bernardi_etal.2014} in order to form forecast combinations of the VaR levels delivered by the SSM, conditional on model's past out--of--sample performance as in \cite{samuels_sekkel.2011} and \cite{samuels_etal.2013}. For further information about the application of the model averaging methodology the reader is referred to \cite{bernardi_etal.2014} where the methodology is extensively discussed.\newline
%%%
The \proglang{R} \citep{R.2013} package MCS here developed provides an integrated environment for the comparison of alternative models or model's specifications within the same family using the Model Confidence Set (MCS) procedure introduced by \cite{hansen_etal.2011}. We believe the main feature lies in the tools that the package provides for specifying different loss functions and the iterative model selection.
%The package is available from the Comprehensive R Archive Network at \url{http://CRAN.R-project.org/package=MCS}.
Throughout the paper, we only assume the reader is familiar with the basic functions and concepts of the \proglang{R} programming language.\newline
%%%
The layout of the paper is as follows. In Section \ref{sec:mcs_procedure} we present the \cite{hansen_etal.2011}'s Model Confidence Set procedure detailing the alternative specifications of the test statistics.
Section \ref{sec:model_specifications} details about the ARCH specifications and their maximum likelihood estimation using the \proglang{R} package \code{rugarch} developed by \cite{ghalanos.2014}. In Section \ref{sec:mcs_package}, hinging on the ARCH--type models described in Section \ref{sec:model_specifications}, we discuss how the package is used to determine the optimal superior set of models (SSM) for the purposes of the user. Section \ref{sec:application} covers the empirical application which aims at illustrating how the procedure in practically implemented to compare several alternative specifications of ARCH--type models. Finally, Section \ref{sec:conclusion} concludes the paper.
%
%:::::::::::::::::::::::::::::::::::::::::::::::::::::::::::::::::::::
% SECTION: MCS PROCEDURE
%:::::::::::::::::::::::::::::::::::::::::::::::::::::::::::::::::::::
\section{The Model Confidence Set procedure}
\label{sec:mcs_procedure}
%:::::::::::::::::::::::::::::::::::::::::::::::::::::::::::::::::::::
%
The availability of several alternative model specifications being able to adequately describe the unobserved data generating process (DGP) opens the question of selecting the \qmo best fitting model\qmcsp according to a given optimality criterion. This paper deals with the MCS procedure developed by \cite{hansen_etal.2011}. The Hansen's procedure consists of a sequence of statistic tests which permits to construct a set of \qmo superior\qmcsp models, the \qmo Superior Set Models\qmcsp (SSM), where the null hypothesis of equal predictive ability (EPA) is not rejected at a certain confidence level. The EPA statistic tests is calculated for an arbitrary loss function that satisfies general weak stationarity conditions, which essentially means that we could test models on various aspects, e.g. punctual forecasts as in \cite{hansen_lunde.2005} or the in--sample goodness of fit as in \cite{hansen_etal.2011}. Formally, let $Y_t$ the observation at time $t$ and $\hat{Y}_{i,t}$ the output of model $i$ at time $t$, the loss function $\ell_{i,t}$ associated to the $i$--th model is defined as
%%%
\begin{equation}
\ell_{i,t}=\ell\left(Y_t,\hat{Y}_{i,t}\right),
\label{eq:loss}
\end{equation}
and measures the difference between the output $\hat{Y}_{i,t}$ and the \textit{a posteriori} realisation $Y_t$. As an example of loss function, \cite{bernardi_etal.2014} consider the asymmetric VaR loss function of \cite{gonzalez_etal.2004} to compare the ability of different GARCH specification to predict extreme loss. Let $\mathrm{VaR}_{t}^{\tau}$ the $\tau$--level predicted VaR at time $t$, given information up to time $t-1$, $\mathcal{F}_{t-1}$, the asymmetric VaR loss function is defined as
\begin{eqnarray}
\ell\left(y_{t},{\rm VaR}_{t}^{\tau}\right)=\left(\tau-d_{t}^{\tau}\right)\left(y_{t}-\mathrm{VaR}_{t}^{\tau}\right),
\end{eqnarray}
where $d_{t}^{\tau}=\bbone\left(y_{t}<\mathrm{VaR}_{t}^{\tau}\right)$ is the $\tau$--quantile loss function, and represents the natural candidate to backtest quantile--based risk measures since it penalises more heavily observations below the $\tau$--th quantile level, i.e. $y_t<\mathrm{VaR}^{\tau}_t$. More details about the loss functions specification can be found in \cite{hansen_lunde.2005} and in the following Section \ref{sec:loss_functions}.\newline
%%%
We now briefly describe how the MCS procedure is implemented. The procedure starts from an initial set of models $\mathrm{\hat{M}}^{0}$ of dimension $m$ encompassing all the model specifications described in Section \ref{sec:model_specifications}, and delivers, for a given confidence level $1-\alpha$, a smaller set, the superior set of models, SSM, $\mathrm{\hat{M}}_{1-\alpha}^{*}$ of dimension $m^*\leq m$. The best scenario is when the final set consists of a single mode, i.e. $m^*=1$. Formally, let $d_{ij,t}$ denotes the loss differential between models $i$ and $j$:
\begin{eqnarray}
d_{ij,t}=\ell_{i,t}-\ell_{j,t},\quad i,j=1,\dots,m,\quad t=1,\dots,n,
\end{eqnarray}
and let
\begin{eqnarray}
d_{i\cdot,t}=\left(m-1\right)^{-1}\sum_{j\in\mathrm{M}}d_{ij,t}\quad i=1,\dots,m,
\end{eqnarray}
be the simple loss of model $i$ relative to any other model $j$ at time $t$. The EPA hypothesis for a given set of models ${\rm M}$ can be formulated in two alternative ways:
\begin{eqnarray}
\mathrm{H}_{0,{\rm M}} & : & c_{ij}=0,\qquad\mathrm{for~all}\quad i,j=1,2,\dots,m\nonumber\\
\mathrm{H}_{{\rm A},{\rm M}} & : & c_{ij}\neq 0,\qquad\mathrm{for~some}\quad i,j=1,\dots,m,
\label{eq:EPA1}
\end{eqnarray}
or
\begin{eqnarray}
\mathrm{H}_{0,{\rm M}} & : & c_{i\cdot}=0,\qquad\mathrm{for~all}\quad i=1,2,\dots,m\nonumber\\
\mathrm{H}_{{\rm A},{\rm M}} & : & c_{i\cdot}\neq 0,\qquad\mathrm{for~some}\quad i=1,\dots,m,
\label{eq:EPA2}
\end{eqnarray}
where $c_{ij}=\mathbb{E}\left(d_{ij}\right)$ and $c_{i\cdot}=\mathbb{E}\left(d_{i\cdot}\right)$ are assumed to be finite and not time dependent.
%%%
According to \citet{hansen_etal.2011}, in order to test the two hypothesis above, the following two statistics are constructed:
\begin{eqnarray}
t_{ij}=\frac{\bar d_{ij}}{\sqrt{\widehat{\rm var}\left(\bar d_{ij}\right)}}\qquad
%\mathrm{T}_{\rm R,M}=\max_{i,j \in\mathrm{M}}
\mathrm{and}\quad
t_{i\cdot}=\frac{\bar d_{i,\cdot}}{\sqrt{\widehat{\rm var}\left(\bar d_{i,\cdot}\right)}}\quad\mathrm{for}\quad i,j\in\mathrm{M},
%\mathrm{T}_{\max ,M}\max_{i,j \in \mathrm{M}}
\label{eq:hansen_test_statistics}
\end{eqnarray}
where $\bar{d}_{i,\cdot}=\left(m-1\right)^{-1}\sum_{j\in\mathrm{M}}\bar d_{ij}$ is the simple loss of the $i$--th model relative to the averages losses across models in the set $\mathrm{M}$, and $\bar{d}_{ij}=m^{-1}\sum_{t=1}^m d_{ij,t}$ measures the relative sample loss between the $i$--th and $j$--th models, while $\widehat{\rm var}\left(\bar d_{i,\cdot}\right)$ and $\widehat{\rm var}\left(\bar d_{ij}\right)$ are bootstrapped estimates of ${\rm var}\left(\bar d_{i,\cdot}\right)$  and ${\rm var}\left(\bar d_{ij}\right)$, respectively. According to \cite{hansen_etal.2011}, to calculate the bootstrapped variances $\widehat{\rm var}\left(\bar d_{i,\cdot}\right)$, we perform a block--bootstrap procedure of 5000 resamples, where the block length $p$ is the max number of significants parameters obtained by fitting an AR$(p)$ process on all the $d_{ij}$ terms. However, in the \code{MCS} package we allow the user to provide an arbitrary block length $p$. Details about the implemented bootstrap procedure can be found in \cite{white.2000}, \cite{kilian.1999}, \cite{clark_mccracken.2001}, \cite{hansen_etal.2003}, \cite{hansen_lunde.2005}, \cite{hansen_etal.2011} and \cite{bernardi_etal.2014}.
%%%
The first \textit{t}--statistic $t_{ij}$ is used in the well know test for comparing two forecasts; see e.g. \citet{diebold_mariano.2002} and \citet{west.1996}, while the second one is used in \citet{hansen_etal.2003,hansen_reinhard.2005,hansen_etal.2011}.
%%%
As discussed in \cite{hansen_etal.2011} the two EPA null hypothesis presented in equation \eqref{eq:EPA1} and \eqref{eq:EPA2} map naturally into the two test statistics:
\begin{eqnarray}
\mathrm{T}_{\rm R,M}=\max_{i,j \in\mathrm{M}}\mid t_{ij}\mid\quad\mathrm{and}\quad\mathrm{T}_{\max ,\rm M}=\max_{i \in \mathrm{M}}t_{i\cdot},
\label{eq:epa_test_stat}
\end{eqnarray}
where $t_{ij}$ and $t_{i.}$ are defined in equation \eqref{eq:hansen_test_statistics}. The test statistics defined in equation \eqref{eq:epa_test_stat} can be used in order to test the two hypothesis \eqref{eq:EPA1} and \eqref{eq:EPA2}, respectively. Since the asymptotic distributions of the two tests statistic is nonstandard, the relevant distributions under the null hypothesis is estimated using a bootstrap procedure similar to that used to estimate ${\rm var}\left(\bar d_{i,\cdot}\right)$ and ${\rm var}\left(\bar d_{ij}\right)$, see for details \citet{white.2000,hansen_etal.2003,hansen_reinhard.2005,kilian.1999,clark_mccracken.2001}.\newline
\indent As said in the Introduction, the MCS procedure consists on a sequential testing procedure, which eliminates at each step the worst model, until the hypothesis of equal predictive ability (EPA) is accepted for all the models belonging to the SSM. The choice of the worst model to be eliminated has been made using an elimination rule that is coherent with the statistic test defined in equation \eqref{eq:hansen_test_statistics} which are
\begin{equation}
e_{\max,\rm M}=\arg\max_{i\in\rm M}\frac{\bar{d}_{i,\cdot}}{\widehat{\rm var}\left(\bar d_{i,\cdot}\right)},\qquad
e_{\rm R,M}=\arg\max_{i}\left\{\sup_{j \in \rm M}\frac{\bar{d}_{ij}}{\sqrt{\widehat{\rm var}\left(\bar d_{ij}\right)}}\right\},
\label{eq:hansen_test_elimination_rule}
\end{equation}
respectively.\newline\newline
Summarazing, the MCS procedure to obtain the SSM, consists of the following steps:
\begin{itemize}
\item[1.] set $\mathrm{M}=\mathrm{M}_{0}$
\item[2.] test for EPA--hypothesis: if EPA is accepted terminate the algorithm and set $\mathrm{M}_{1-\alpha}^{*}=\mathrm{M}$, otherwise use the elimination rules defined in equations \eqref{eq:hansen_test_elimination_rule} to determine the worst model.
\item[3.] remove the worst model, and go to step 2.
\end{itemize}
%
%In the next sections, we apply the MCS procedure using the $\mathrm{T}_{\max ,\rm M}$ statistic tests and we verify that, in our case, results are independent to that choice.
%
Since the Hansen's procedure usually delivers a SSM $\hat{\sM}_{1-\alpha}^*$ which contains a large number of models, in the next sections, we also describe how to implement a procedure that combines the VaRs forecasts delivered by the MCS procedure.
%
%:::::::::::::::::::::::::::::::::::::::::::::::::::::::::::::::::::::
% SUBSECTION: MODEL SPECIFICATIONS
%:::::::::::::::::::::::::::::::::::::::::::::::::::::::::::::::::::::
\section{Model specifications}
\label{sec:model_specifications}
%:::::::::::::::::::::::::::::::::::::::::::::::::::::::::::::::::::::
%
\noindent In the empirical illustration we apply the MCS procedure detailed in the previous Section \ref{sec:mcs_procedure} to compare the VaR forecasts obtained by fitting a list of popular autoregressive conditional heteroskedastic models introduced by \cite{engle.1982} and \cite{bollerslev.1986}. Here, we chose the ARCH--type models, because of their popularity, but the procedure could be implemented for any kind of model without any further complications. To account for the stylised facts about financial returns we consider several specifications differing for the conditional mean, the volatility dynamics and the distributions of the error term. Formally, let $y_t$ the logarithmic return at time $t$, we consider the following general AR(1)--GARCH--in mean specification originally proposed by \citet{engle_etal.1987}
\begin{align}
y_t&=\mu+\lambda\sigma_t^2 + \phi y_{t-1}+\varepsilon_t,\qquad\varepsilon_t=\sigma_t\zeta_t,\qquad\zeta_t\sim\mathcal{D}\left(0,1\right)\nonumber\\
%%%
\sigma_t^2&=h\left(\sigma_{t-1},\varepsilon_{t-1},\btheta_\sigma\mid\mathcal{F}_{t-1}\right),\nonumber
\end{align}
where $\mathcal{F}_t$ is the information set up to time $t$, $\zeta_t$ is a sequence of independently and identically distributed random variables with general standardized distribution $\mathcal{D}$, $\sigma_t$ is the conditional standard deviation of $y_t$ and $\phi$ is the autoregressive parameter assumed to be $\vert\phi\vert<1$ to preserve stationarity. The risk--premium parameter $\lambda$ is set equal to zero if the \qmo in mean\qmcsp specification is omitted, otherwise it is jointly estimated with the other parameters. Finally, the function $h\left(\cdot\right)$ refers to one of the ARCH--type dynamics reported below, where the vector $\btheta_\sigma$ contains all the conditional variance dynamic parameters.\newline\newline
%%%
Among the most popular distributions that econometricians usually choose to model the error term $\varepsilon_t$ we consider: the Gaussian $\mathcal{N}\left(0,1\right)$, the Student--t $\mathcal{T}_{\nu}\left(0,\frac{\nu-2}{\nu}\right)$, with $\nu$ degrees of freedom, the Generalised Error distribution $\mathcal{GED}\left(0,1,\kappa\right)$, where $\kappa$ is the shape parameter, and their asymmetric counterparts, the Skew--Normal, the Skew Student--t and the Skew--GED, obtained by applying the skewing mechanism of \cite{fernandez_steel.1998}. %or the arguments presented in \cite{azzalini.1985} and \cite{azzalini_capitanio.2003}.\newline\newline
Furhermore, we consider the Johnson's reparametrised SU distribution of \cite{rigby_stasinopoulos.2005} and Generalized Hyperbolic introduced by \citet{barndorff.1977}.\newline\newline%as a subclass of the generalised hyperbolic distribution.\newline\newline
ARCH--type models are flexible and powerful tools for conditional volatility modelling, because they are able to consider the volatility clustering phenomena and other established stylised facts such as excess of kurtosis and asymmetry of the unconditional returns. Models belonging to this class, was principally proposed in order to describe the time--varying nature of the conditional volatility that characterises financial assets. They have also become one of the most used tool for researchers and practitioners dealing with financial market exposure. The simplest conditional volatility dynamics we consider is the GARCH(p,q) specification introduced by \cite{bollerslev.1986}
\begin{equation}
\sigma_t^2=\omega+\sum_{i=1}^p\alpha\varepsilon_{t-i-1}^2+\sum_{j=1}^q\beta\sigma_{t-j-1}^2,
\label{eq:garch_pq_spec}
\end{equation}
where $\omega>0$ and $0\leq\alpha_i<1,\forall i=1,2,\dots,p$ and $0\leq\beta_j<1,\forall j=1,2,\dots,q$ with $P\equiv\sum_{i=1}^p\alpha+\sum_{j=1}^q\beta<1$ to preserve weak ergodic stationarity of the conditional variance. Sometimes it is possible to observe high persistence in financial time series volatility, i.e. it is possible to observe series for which $P\approx1$. To account for this scenario the IGARCH(p,q) specification, where the persistence parameters is imposed to be exactly 1, i.e $P=1$, has been proposed by \cite{engle_bollerslev.1986}. Despite their  popularity, the GARCH and the IGARCH specifications are not able to account for returns exhibiting higher volatility after negative shocks than after positive ones as theorised by the \qmo leverage effect\qmcsp of \cite{black.1976}. Consequently, in the financial econometric literature several ARCH models have been proposed. The EGARCH(p,q) model of \cite{nelson.1991}, for example, assumes that the conditional volatility dynamics follows
\begin{equation}
\log\left(\sigma_t^2\right)=\omega+\sum_{i=1}^p\left[\alpha_i\zeta_{t-i}+\gamma_i\left(\vert\zeta_{t-i}\vert-\mathbb{E}\vert\zeta_{t-i}\vert\right)\right]+\sum_{j=1}^q\beta_j\log\left(\sigma_{t-j}^2\right),
\label{eq:egarch_pq_spec}
\end{equation}
where the asymmetric response is introduced through the $\gamma_i$ parameters: for $\gamma_i<0$ negative shocks will obviously have a bigger impact on future volatility than positive shocks of the same magnitude. Note that $\alpha_i,\beta_j,\gamma_i$ could assume any value and no positivity constraints are required. For the EGARCH(p,q) specification the persistence parameter $P$ is equal to $P=\sum_{j=1}^q\beta_j$. Another widely used asymmetric GARCH model is the GJR--GARCH(p,q) specification of \citet{glosten_etal.1993}, which accounts for the leverage effect simply by including a dummy variable which discriminates positive and negative lagged shocks in the following way
\begin{equation}
\sigma_t^2=\omega+\sum_{i=1}^p\left(\alpha_i +\gamma_i\mathbb{I}_{\{\varepsilon_{t-i}<0\}}\right)\varepsilon_{t-i}^2+\sum_{j=1}^q\beta_j\sigma_{t-j}^2,
\label{eq:gjr_garch_pq_spec}
\end{equation}
where $\mathbb{I}_{\{\varepsilon_{t-i}<0\}}$ assumes value one if $\varepsilon_{t-i}<0$ for $i=1,2,\dots,p$ and zero otherwise. Because of the presence of the indicator function, the persistence of the GJR--GARCH specification crucially depends on the asymmetry of the conditional distribution used to model the error term $\varepsilon_t$
%%%
\begin{equation}
P=\sum_{i=1}^p \alpha_i+\sum_{j=1}^q \beta_j+\sum_{i=1}^p \gamma_i\mathbb{P}\left(\varepsilon_{t-i}\leq 0\right),
\end{equation}
%%%
where $\mathbb{P}\left(\varepsilon_{t-i}\leq 0\right)$ denotes the probability of observing negative shocks and $\omega>0$, $\alpha_i\ge0$, for $i=1,2,\dots,p$, $\beta_j\ge0$, for $j=1,2,\dots,q$, and the additional constraint $\alpha_i+\gamma_i\ge0$ for $i=1,2\dots,p$ is imposed to preserve the positiveness of the conditional variance. Finally, the Asymmetric--Power--ARCH(p,q) (APARCH, henceforth) of \cite{ding_etal.1993} imposes the following dynamic to the conditional variance
\begin{equation}
\sigma_t^{(\delta)}=\omega+\sum_{i=1}^p\alpha_i \left(\vert\varepsilon_{t-i}\vert-\gamma_i\varepsilon_{t-i}\right)^{\delta}+\sum_{j=1}^q\beta_j\sigma_{t-j}^{(\delta)},
\label{eq:aparch_pq_spec}
\end{equation}
where $x^{\left(\delta\right)}=\frac{x^\delta-1}{\delta}$ is the Box--Cox transformation of \citet{box_cox.1964}. The parameters restrictions to ensure the positiveness of the conditional variance are $\omega>0$, $\delta\ge0$, $0\le\gamma_i\le0$ for $i=1,\dots,p$ and the usual condition $\alpha_i\geq0$, and $\beta_j\ge0$, for $i,j=1,2,\dots,\max{\{p,q\}}$. As for the GJR--GARCH specification the persistence strongly depends to the probability density function chosen for the innovation term $\zeta_t$
\begin{equation}
P=\sum_{i=1}^p\alpha_i\kappa_i + \sum_{j=1}^q\beta_j,
\end{equation}
where $\kappa_i=\mathbb{E}\left[\vert\zeta\vert-\gamma_i\zeta\right]^\delta$, for $i=1,\dots,q.$
%%%%
The APARCH specification results in a very flexible model that nests several of the most popular univariate ARCH parameterisations, such as
\begin{itemize}
%%%
\item the GARCH(p,q) of \citet{bollerslev.1986} for $\delta=0$ and $\gamma_i=0$, for $i=1,2,\dots,p$;
%%%
\item the Absolute--Value--GARCH (AVARCH, henceforth) specification for $\delta=1$ and $\gamma_i=0$ for $i=1,2,\dots,p$, proposed by \citet{taylor.1986} and \citet{schwert.1990} to mitigates the influence of large, in an absolute sense, shocks with respect to the traditional GARCH specification;
%%%
\item the GJR--GARCH model of \citet{glosten_etal.1993} for $\delta=2$ and $0\le\gamma_i\le1$ for $i=1,2,\dots,p$;
%%%
\item the Threshold GARCH (TGARCH, henceforth) of \citet{zakoian.1994} for $\delta=1$, which allows different reactions of the volatility to different signs of the lagged errors;
%%%
\item the Nonlinear GARCH (NGARCH, henceforth) of \citet{higgins_bera.1992} for $\gamma_i=0$ for $i=1,2,\dots,p$ and $\beta_j=0$ for $j=1,2,\dots,q$.
%%%
%\item the Log--ARCH model of \citet{geweke.1986} and \citet{pantula.1986} for $\delta\to0$, proposed to remove the parameter restrictions needed to ensure the positiveness of the conditional variance. The Log--ARCH is not usually employed because the conditional variance dynamic is not defined for $\varepsilon_t=0$.
\end{itemize}
%%%
Another interesting specification is the Component--GARCH(p,q) (CGARCH, henceforth) of \citet{engle_lee.1993} which decomposes the conditional variance into a permanent and transitory component in a straightforward way
%%%
\begin{align}
\sigma_t^2&=\xi_t+\sum_{i=1}^p\alpha_i\left( \epsilon_{t-i}^2-\xi_{t-i}\right)+\sum_{j=1}^q\beta_j\left(\sigma_{t-j}^2-\xi_{t-j}\right)\nonumber\\
%%%
\xi_t&=\omega+\rho \xi_{t-1}+\eta\left(\epsilon_{t-1}^2-\sigma_{t-1}^2\right),
%%%
\label{eq:csgarch_pq_spec}
\end{align}
%%%
where to ensure the stationarity of the process at the usual $\sum_{i=1}^p\alpha_i + \sum_{j=1}^q\beta_j<1$ condition the $\rho<1$ constrain must be added. Further parameters restrictions for the positiveness of the conditional variance are given in \cite{engle_lee.1993}. This solution is usually employed because it permits to investigate the long and short--run movements of volatility.
The considered conditional volatility models are a minimal part of the huge number of specifications available in the financial econometric literature. We chose these models because of their heterogeneity, since each of them focuses on a different kind of stylised fact. Moreover, even if they could seem very similar, the way in which they account for the stylised fact changes. For a very extensive and up to date survey on GARCH models we will refer the reader to the works of \cite{bollerslev.2008}, \cite{terasvirta.2009}, \cite{bauwens_etal.2006}, \cite{silvennoinen_terasvirta.2009} and the recent book of \cite{francq_zakoian.2011}. To estimate model parameters we consider the maximum likelihood approach, see e.g. \cite{francq_zakoian.2011}.
%
%:::::::::::::::::::::::::::::::::::::::::::::::::::::::::::::::::::::
% SUBSECTION: GARCH R
%:::::::::::::::::::::::::::::::::::::::::::::::::::::::::::::::::::::
\subsection{GARCH model estimation and forecast in R}
\label{sec:garch_estimation_R}
%:::::::::::::::::::::::::::::::::::::::::::::::::::::::::::::::::::::
%
Within the R environment a wide range of statistical packages are available in order to deal with GARCH models estimation and forecast such as the {\tt fGarch} package of \cite{fgarch}. Here, in order to estimate the competing GARCH models we use some functions belonging to the library \textit{rugarch} developed by \cite{ghalanos.2014}. Of course, the following treatment is only useful for illustrative purposes on the use of the \texttt{MCS} package, and readers are free to choose a different package to estimate models as well as their handwritten functions. Before starting the MCS procedure, it is necessary to define the set of competing GARCH models ${\rm M}_0$. This can be done using the the {\tt ugarchspec()} function that permits to specify a variety of GARCH models such as those described in the previous Section \ref{sec:model_specifications}. For example, the following portion of code
%%%
\begin{CodeChunk}
\begin{CodeInput}
R> library(rugarch)
R> spec <- ugarchspec(mean.model = list(armaOrder = c(0, 0)),
                      variance.model = list(model = "sGARCH",
                      garchOrder = c(1, 1))
\end{CodeInput}
\end{CodeChunk}
creates an {\tt uGARCHspec} object \qmo {\tt spec}\qmcsp which defines a GARCH(1,1) model with gaussian innovations (the chosen conditional distribution can be changed by using the \textit{distribution.model} argument). The object {\tt spec} can be subsequently used into the {\tt ugarchfit()} function in order to estimate the model on a given time series. For example
%%%
\begin{CodeChunk}
\begin{CodeInput}
R> library(MCS)
R> data(STOXXIndexesRet)
R> ret <- STOXXIndexesRet[,"SXA1E"]
R> fit <- ugarchfit(spec = spec, data = ret)
\end{CodeInput}
\end{CodeChunk}
creates an {\tt uGARCHfit} object \qmo{\tt fit}\qmcsp containing parameter estimates of the STOXX North America 600 index (\qmo SXA1E\qmcsp) as well as several additional informations such as the Information Criteria, tests on standardised residuals, and several ARCH LM tests. Prediction using GARCH models can be easily performed using the {\tt ugarchforecast()} function which takes as an argument the output of the fitting procedure {\tt uGARCHfit} or alternatively the {\tt uGARCHspec} object. The one step ahead forecasts can be easily obtained using the following routine
%%%
\begin{CodeChunk}
\begin{CodeInput}
R> OneStepForc <- ugarchforecast(fitORspec = fit, n.ahead = 1)
\end{CodeInput}
\end{CodeChunk}
%%%
which reports an {\tt uGARCHforecast}, object \qmo{\tt OneStepForc}\qmc. The \texttt{rugarch} package also includes the {\tt ugarchroll()} function which permits to construct a series of one step ahead rolling forecasts allowing also the user to define a \qmo refit window\qmcsp and the length of the forecast series. For example, a rolling forecast series of length $2000$, using a refit window of 5 observations can be computed as follow
%%%
\begin{CodeChunk}
\begin{CodeInput}
R> roll <- ugarchroll(spec = spec, data = ret, forecast.length = 2000,
                      refit.every = 5)
\end{CodeInput}
\end{CodeChunk}
which reports an {\tt uGARCHroll} object {\tt roll}. Finally, a variety of methods are present in the \textit{rugarch} package in order to deal with \qmo {\tt uGARCHspec}\qmcsp, \qmo {\tt uGARCHfit}\qmcsp, \qmo {\tt uGARCHforecast}\qmcsp and \qmo {\tt uGARCHroll}\qmcsp objects. For more informations see \citet{ghalanos.2014} or consult the help() in {\tt R}.
%
%:::::::::::::::::::::::::::::::::::::::::::::::::::::::::::::::::::::
% SUBSECTION: GARCH R
%:::::::::::::::::::::::::::::::::::::::::::::::::::::::::::::::::::::
\section{Using the package}
\label{sec:mcs_package}
%:::::::::::::::::::::::::::::::::::::::::::::::::::::::::::::::::::::
%
As described in Section \ref{sec:mcs_procedure} the MCS procedure is used to compare different models under an user defined loss function. The loss function measures the \qmo performance\qmcsp of the competing models at a each time point $t$ in the evaluating period. Suppose now to have $m$ competing models and an evaluating period of length $n$, then using the defined loss function it is possible to construct a loss matrix of dimension $\left(m\times n\right)$. Then the {\tt MCSprocedure()} function can be used to construct the set of superior models outlined in Section \ref{sec:mcs_procedure}.
%
%:::::::::::::::::::::::::::::::::::::::::::::::::::::::::::::::::::::
% SUBSECTION: GARCH R
%:::::::::::::::::::::::::::::::::::::::::::::::::::::::::::::::::::::
\subsection{Comparing GARCH models using MCS}
\label{sec:mcs_garch_R}
%:::::::::::::::::::::::::::::::::::::::::::::::::::::::::::::::::::::
%
The MCS procedure can be used to compare models under various aspects, for example with respect to their ability to predict the future volatility, or the future returns conditional to actual and past information. Suppose instead we are interested to compare the VaR forecasts delivered by different GARCH models. As previously explained, the {\tt ugarchroll()} function can be used to obtain a one step ahead rolling forecast series of a specified GARCH model. Furthermore, the {\tt as.data.frame()} method permits to extract from the {\tt uGARCHroll} object the VaR forecasts series at both the 1\% and 5\% confidence levels. For example, if we want to compare five different GARCH specifications, such as
%%%
\begin{itemize}
\item the GARCH(1,1) of \cite{bollerslev.1986}, \qmo{\tt sGARCH}\qmcsp in the {\tt rugarch} package;
\item EGARCH(1,1) of \cite{nelson.1991}, \qmo{\tt eGARCH}\qmcsp in the {\tt rugarch} package;
\item GJRGARCH(1,1) of \cite{glosten_etal.1993}, \qmo{\tt gjrGARCH}\qmcsp in the {\tt rugarch} package;
\item APARCH(1,1) of \cite{ding_etal.1993}, \qmo{\tt apARCH}\qmcsp in the {\tt rugarch} package;
\item CGARCH(1,1) of \citet{engle_lee.1993}, \qmo{\tt csGARCH}\qmcsp in the {\tt rugarch} package;
\end{itemize}
%%%
we can simply define the five GARCH specifications combined with the distributions presented in the previous section for the innovation terms using the {\tt ugarchspec()} function as follows
%%%
\begin{CodeChunk}
\begin{CodeInput}
R> models <- c("sGARCH", "eGARCH", "gjrGARCH", "apARCH", "csGARCH")
R> distributions <- c("norm", "std", "ged", "snorm", "sstd", "sged", "jsu", "ghyp")
R> spec.comp <- list()
R> for( m in models ) {
     for( d in distributions ) {
   spec.comp[[paste( m, d, sep = "-" )]] <-
      ugarchspec(mean.model = list(armaOrder = c(0, 0)),
      variance.model = list(model = m, garchOrder = c(1, 1)),
      distributtion.model=d)
                                }
                      }
R> specifications <- names( spec.comp )
\end{CodeInput}
\end{CodeChunk}
%%%
In this way we have defined a list with the 40 combinations of GARCH specifications and innovation term distributions, and we can perform the rolling forecast of length $2000$ using a \qmo refit window\qmcsp of $200$ observations
%%%
\begin{CodeChunk}
\begin{CodeInput}
R> roll.comp <- list()
R> for( s in specifications ){
   roll.comp[[s]] <-  ugarchroll(spec = spec.comp[[s]], data = ret,
                           forecast.length = 2000, refit.every = 200)
                      }
\end{CodeInput}
\end{CodeChunk}
and, using the {\tt as.data.frame()} method, extract the VaR forecast series at the confidence level $\tau=1\%$
\begin{CodeChunk}
\begin{CodeInput}
R> VaR.comp=list()
R> for( s in specifications ) {
   VaR.comp[[s]] <-  as.data.frame(roll.comp[[s]], which = "VaR")[, 1]
   }
\end{CodeInput}
\end{CodeChunk}
Now it is possible to calculate the loss associated to each model at each time. Here, we consider the asymmetric VaR loss function of \cite{gonzalez_etal.2004} and considered also by \cite{bernardi_etal.2014}, which is implemented in the {\tt MCS} package through the {\tt LossVaR()} function, more details of the available loss functions are reported in Section \ref{sec:loss_functions}
%%%
\begin{CodeChunk}
\begin{CodeInput}
R> Loss <- do.call(cbind,lapply(specifications,
                   function(s) LossVaR(tau=0.01, realized=tail(ret, 2000)/100,
                   evaluated=VaR.comp[[s]]/100)))
R> colnames(Loss) <- specifications
\end{CodeInput}
\end{CodeChunk}
The object \qmo{\tt Loss}\qmcsp is a matrix of dimension $\left(2000\times 40\right)$ that contains the VaR losses associated to each of the different GARCH specifications we considered in this example. The object \qmo{\tt Loss}\qmcsp is also included in the MCS packages and can be easily loaded using {\tt data(Loss)}, since it represents the main input of the {\tt MCSprocedure()} function described in Section \ref{sec:ssm_garch_R}. Nevertheless, in the next section we describe some alternative loss functions which are particularly suitable for volatility forecast assessment as well as to forecast future observations conditional to past information.
%
%:::::::::::::::::::::::::::::::::::::::::::::::::::::::::::::::::::::
% SUBSECTION: APPLICATION OF VAR COMBINATION
%:::::::::::::::::::::::::::::::::::::::::::::::::::::::::::::::::::::
\subsection{Loss functions}
\label{sec:loss_functions}
%:::::::::::::::::::::::::::::::::::::::::::::::::::::::::::::::::::::
%
As previously discussed the MCS procedure is able to discriminate models under a user defined loss function. The choice of the loss function is somewhat arbitrary, and crucially depends on the nature of the competing models and the scope of their usage. For more considerations about the choice of the loss function for model comparison purposes we refer to \cite{hansen_lunde.2005}, \cite{bollerslev_etal.1994}, \cite{diebold_lopez.1996} and \cite{lopez.2001}. In what follows, we report the loss functions available within the \code{MCS} package. However, since the \code{MCSprocedure()} function described in the previous section accepts a pre--defined loss matrix, named \qmo\code{Loss}\qmc, the user is free to define and use its own loss function.
%%%
Three different loss functions are freely available within the \code{MCS} package:
\begin{itemize}
\item[1.] the \code{LossVaR()} that can be used to check the performances associated to VaR (or more generally quantile) forecasts;
\item[2.] the \code{LossVol()} for volatility forecasts assessment;
\item[3.] the \code{LossLevel()} that can be used instead for level forecasts, as the punctual mean forecast of a regression model.
\end{itemize}
These loss functions accept three common arguments. The first two arguments are \code{realized} that consists of a vector of realised observations (i.e. the ones that a model hopes to accurately forecast or describe), and \code{evaluated} which is a vector or a matrix of models output. Note that we decided to call the second argument of those functions \qmo \code{evaluated}\qmcsp instead of \qmo \code{forecasted}\qmcsp since the MCS procedure is more general than simply a procedure suited for forecasts evaluation. In fact, as reported by \cite{hansen_etal.2011} the MCS procedure also adapts to in sample studies. The third argument \code{which} instead is function dependent. The available choices and other function specific arguments are reported below.
%%%
\begin{itemize}
\item For \code{LossVaR()} only \code{which = "asymmetricLoss"} is available. This coincides with the asymmetric VaR loss function of \cite{gonzalez_etal.2004} which is particularly suited to assess quantile risk measures, such as the VaR, since it penalises more heavily observations below the $\tau$--th quantile level, i.e. $y_t<\mathrm{VaR}_t^{\tau}$. The asymmetric loss function is defined as
\begin{eqnarray}
\ell\left(y_{t},{\rm VaR}_{t}^{\tau}\right)=\left(\tau-d_{t}^{\tau}\right)\left(y_{t}-\mathrm{VaR}_{t}^{\tau}\right),
\label{eq:asy_loss_gonzalez_rivera}
\end{eqnarray}
where $d_{t}^{\tau}=\bbone\left(y_{t}<\mathrm{VaR}_{t}^{\tau}\right)$ is the $\tau$--quantile loss function.
%%%
Further arguments are \code{tau}, which represents the VaR confidence level and \code{type} with possible choices \code{"normal"} and \code{"differentiable"}. The \code{type} argument permits to discriminate between the normal and the differentiable versions of the loss function: \code{"normal"} permits the specification of the loss function of \cite{gonzalez_etal.2004} defined in equation \eqref{eq:asy_loss_gonzalez_rivera} while \code{"differentiable"} considers the following loss function
%%%
\begin{equation}
\ell\left(r_{t},\mathrm{VaR}_{t}^\tau\right)=\left(\tau-m_{\delta}\left(r_{t},\mathrm{VaR}_{t}^\tau\right)\right)\left(r_{t}-\mathrm{VaR}_{t}^\tau\right),
\end{equation}
%%%
where $m_\delta\left(a,b\right)=\left[1+\exp{\{\delta\left(a-b\right)\}}\right]^{-1}$. Note that the $\delta$ parameter, controlling the function smoothness, can be chosen by the \code{delta} argument in the \code{LossVaR()} function and it is set equal to $25$ by default.
%%%
\item For \code{LossVol()}, the six loss functions reported in \cite{hansen_lunde.2005} are implemented. Note that for this kind of loss functions the \code{realized} and the \code{evaluated} arguments should be some realised volatility measures $\tilde\sigma_{t+1}$ and the punctual volatility forecasts $\hat\sigma_{t+1}$. In this context, we use the term volatility as for the standard deviation $\sigma$. The implemented loss functions are:
%%%
\begin{itemize}
\item[1.] $\mathrm{SE}_{1,t+1}=\left(\tilde\sigma_{t+1}-\hat\sigma_{t+1}\right)^2$, by setting \code{which = "SE1"},
\item[2.] $\mathrm{SE}_{2,t+1}=\left(\tilde\sigma_{t+1}^2-\hat\sigma_{t+1}^2\right)^2$, by setting \code{which = "SE2"},
\item[3.] $\mathrm{QLIKE}_{t+1}=\log{\left(\hat\sigma_{t+1}^2\right)+\tilde\sigma_{t+1}^2 \hat\sigma_{t+1}^{-2}}$, by setting \code{which = "QLIKE"},
\item[4.] $\mathrm{R}^2\mathrm{LOG}_{t+1}=\left[\log\left(\tilde\sigma_{t+1}^2 \hat\sigma_{t+1}^{-2}\right)\right]^2$, by setting \code{which = "R2LOG"},
\item[5.] $\mathrm{AE}_{1,t+1}=\vert \tilde\sigma_{t+1} - \hat\sigma_{t+1}\vert$, by setting \code{which = "AE1"},
 \item[6.] $\mathrm{AE}_{2,t+1}=\vert \tilde\sigma_{t+1}^2 - \hat\sigma_{t+1}^2\vert$, by setting \code{which = "AE2"}.
\end{itemize}
%%%
\item For \code{LossLevel()}, the \code{which} argument accepts values: \code{"SE"} and \code{"AE"} that coincide with the squared error and the absolute error.
\end{itemize}
%
%:::::::::::::::::::::::::::::::::::::::::::::::::::::::::::::::::::::
% SUBSECTION: GARCH R
%:::::::::::::::::::::::::::::::::::::::::::::::::::::::::::::::::::::
\subsection{Constructing the Superior Set of Models}
\label{sec:ssm_garch_R}
%:::::::::::::::::::::::::::::::::::::::::::::::::::::::::::::::::::::
%
The function {\tt MCSprocedure()} returns a S4 object of the class \qmo{\tt SSM}\qmc, which has several arguments we now briefly describe here. The main inputs of the function {\tt MCSprocedure()} are
%%%
\begin{itemize}
\item[-] \texttt{Loss}, which must be matrix or something coercible to that (using the as.matrix() function) which contains the loss series for each model to be compared.
\item[-] \texttt{alpha}, that must be a positive scalar in $\left(0,1\right)$ indicating the confidence level of the MCS tests.
\item[-] \texttt{B}, which is an integer indicating the number of bootstrapped samples used to construct the statistic test.
\item[-] \texttt{cluster}, that coincides with a cluster object created by calling makeCluster from the parallel package. By default this is set to NULL but if an appropriate cluster object is submitted, then this will be used for parallel processing.
\item[-] \texttt{statistic}, which is the statistic that should be used to test the EPA at each step of the iteration. Possible choices are \qmo Tmax\qmcsp and \qmo TR\qmc, which coincide with $\mathrm{T}_{\max ,M}$ and $\mathrm{T}_{\rm R,M}$ of Section \ref{sec:mcs_procedure} respectively.
\end{itemize}
For sample purposes, in the MCS package the \qmo Loss\qmcsp dataset is included. This coincide with the \texttt{Loss} matrix obtained in the previous Section and can be loaded using \texttt{data(Loss)}. The construction of the Superior Set of Models can be easily done using the following portion of code:\newline\newline
\begin{CodeChunk}
\begin{CodeInput}
R> library(MCS)
R> data(Loss)
R> SSM <- MCSprocedure(Loss = Loss, alpha = 0.2, B = 5000, statistic = "Tmax")
\end{CodeInput}
\begin{CodeOutput}
R> SSM
------------------------------------------------------------------------------
-                            Superior Set of Models                          -
------------------------------------------------------------------------------
              Rank_M          v_M  MCS_M Rank_R         v_R  MCS_R       Loss
sGARCH-ged         27  0.996797601 0.4486     27  1.46532329 0.2352 0.0003986329
sGARCH-snorm       25  0.896954417 0.5920     26  1.38772567 0.2998 0.0003982803
sGARCH-sstd        26  0.938712179 0.5248     25  1.37978372 0.3166 0.0003977886
sGARCH-sged        20  0.521486029 0.9732     20  1.20336841 0.5346 0.0003956815
sGARCH-jsu         23  0.820277282 0.7052     24  1.35541006 0.3166 0.0003971334
sGARCH-ghyp        22  0.685040892 0.8628     22  1.27425551 0.4478 0.0003964821
eGARCH-ged         30  1.136085890 0.2480     31  1.55570183 0.1574 0.0003994099
eGARCH-sstd        17 -0.004284375 1.0000     17  0.99146387 0.7916 0.0003933537
eGARCH-sged        12 -0.538418981 1.0000     12  0.68337534 0.9802 0.0003910679
eGARCH-jsu         15 -0.132370421 1.0000     15  0.92181102 0.8454 0.0003928127
eGARCH-ghyp        16 -0.107676711 1.0000     16  0.94757871 0.8260 0.0003929190
gjrGARCH-norm      18  0.213742841 1.0000     18  1.10231033 0.6710 0.0003943638
gjrGARCH-std       13 -0.427196927 1.0000     13  0.75538350 0.9588 0.0003916026
gjrGARCH-ged        6 -0.954946367 1.0000      8  0.37895234 1.0000 0.0003891435
gjrGARCH-snorm      1 -1.295629237 1.0000      2  0.04458467 1.0000 0.0003870702
gjrGARCH-sstd       5 -0.962068788 1.0000      5  0.33202562 1.0000 0.0003887004
gjrGARCH-sged       2 -1.252823785 1.0000      1 -0.04527337 1.0000 0.0003867926
gjrGARCH-jsu        7 -0.942033667 1.0000      6  0.33760463 1.0000 0.0003887359
gjrGARCH-ghyp       8 -0.929362792 1.0000      4  0.32763998 1.0000 0.0003886976
apARCH-norm        21  0.535927772 0.9668     21  1.25059782 0.4788 0.0003959991
apARCH-std         19  0.367561226 0.9980     19  1.19805785 0.5490 0.0003949103
apARCH-ged         14 -0.317682324 1.0000     14  0.80503488 0.9304 0.0003920526
apARCH-snorm        4 -0.977717453 1.0000      7  0.33884203 1.0000 0.0003889586
apARCH-sstd         9 -0.628176498 1.0000      9  0.63183157 0.9904 0.0003905724
apARCH-sged         3 -1.022889502 1.0000      3  0.29488276 1.0000 0.0003884919
apARCH-jsu         10 -0.588780764 1.0000     10  0.63897617 0.9892 0.0003906327
apARCH-ghyp        11 -0.582730804 1.0000     11  0.64705812 0.9878 0.0003906581
csGARCH-sstd       31  1.165077099 0.2170     30  1.53265345 0.1782 0.0003992507
csGARCH-sged       24  0.839136228 0.6792     23  1.34145394 0.3166 0.0003972841
csGARCH-jsu        29  1.063487090 0.3454     28  1.47330668 0.2280 0.0003985447
csGARCH-ghyp       28  1.046751336 0.3706     29  1.47854362 0.2244 0.0003984870
------------------------------------------------------------------------------
Details
Number of eliminated models : 9
Statistic : Tmax
Elapsed Time : Time difference of 6.771243 mins
\end{CodeOutput}
\end{CodeChunk}
%
%:::::::::::::::::::::::::::::::::::::::::::::::::::::::::::::::::::::
% SECTION: APPLICATION
%:::::::::::::::::::::::::::::::::::::::::::::::::::::::::::::::::::::
\section{Application}
\label{sec:application}
%:::::::::::::::::::::::::::::::::::::::::::::::::::::::::::::::::::::
%
For the empirical study, a panel of four major worldwide stock markets indices is considered. The four daily stock price indices includes the Asia/Pacific 600 (SXP1E), the North America 600 (SXA1E) and the Europe 600 (SXXP) as well as the Global 1800 (SXW1E). The data are freely available and can be download from the STOXX website \url{http://www.stoxx.com/indices/types/benchmark.html}. The data were obtained over a 23--years time period, from 31 December 1991 to 24 July 2014, for a total of 5874 observations. For each market, the returns are calculated as the logarithmic difference of the daily price index and multiplied by 100
%%%
\begin{equation}
y_t=\left(\log\left(p_t\right)-\log\left(p_{t-1}\right)\right)\times 100,\nonumber
\end{equation}
%%%
where $p_t$ is the closing index value on day $t$.
%%%
To examine the performance of the models to predict extreme VaR levels, the complete dataset of daily returns is divided into two samples: an in--sample period from 1 January 1992 to 06 October 2006, for a total of 3814 observations, and a forecast or validation period, containing the remaining 2000 observations: from 09 October 2006 to 24 July 2014. A rolling window approach is used to produce 1--day ahead forecasts of the 1\% VaR thresholds $\mathrm{VaR}_{t+1}^{0.01}$, for $t=1,2,\dots,1999$ in the forecast samples. To apply the MCS procedure we consider $160$ ARCH--type specifications obtained by combining the models reported in the previous Section \ref{sec:model_specifications} estimated on each of the four international indexes. More precisely, the $160$ model specifications have been obtained by considering all the possible combinations of the $10$ ARCH dynamics, the $8$ conditional distributions and the \textit{in--mean/non in--mean} options as outlined in Section \ref{sec:model_specifications}. Estimated coefficients for each fitted model are not reported but they are available upon request to the second author. VaR estimates are performed by inverting the conditional cumulative density function of the corresponding estimated model. Then, the MCS procedure of \citet{hansen_etal.2011} described in the previous Section \ref{sec:mcs_procedure} is applied to obtain the set of models with superior predictive ability in term of the VaR forecast at the 1\% confidence level.\newline
%%%%
Table \ref{tab:Composition of remaining models} reports the compositions of the Superior Set of Models discriminating by model, distribution and \textit{in--mean} options. The different entries in each column represent the number of models that belong to the SSM at the end of the MCS procedure discriminated by model, distribution and \textit{in--mean} options. From Table \ref{tab:Composition of remaining models} we can observe that for the SXA1E and SXP1E indexes the SSM is quite homogeneous with respect to the dynamics and innovation assumptions. In these case the one step ahead 1\% VaR forecasting performance of the competing models are quite similar, suggesting that for those series the use of complicated nonlinear models are not entirely justified. Moreover, it is interestingly to note that the EGARCH specification is the model most frequently eliminated, probably because the logarithmic specification of the conditional volatility is too much sensitive to the previous volatility changes. Concerning the distribution specifications, we observe that the MCS procedure confirms the common finding that the Gaussian distribution poorly describes the financial time series behaviour. Conversely, looking at the third and the fourth columns, it is possible to asses that for the SXW1E and SXXP indexes the MCS procedure reports more discriminant results. For the SXW1E index, all the specifications that include the GARCH, IGARCH, C--GARCH and N--GARCH dynamics were eliminated while also the GJR--GARCH dynamic is not present in the SSM for the SXXP index. The exclusion of those models suggests that for the SXW1E and SXXP series, more complicated dynamics are necessary to describe the shape of the conditional returns density function.\newline
%%%%
In order to test the benefit of using the MCS procedure technique we report a simple comparison study using two different VaR forecasts. The first VaR forecasts series ($\mathrm{VaR_{avg}}$) is performed using the simple average VaR across all the 160 available models, while the second ($\mathrm{VaR_{Dyn}}$) is performed using the dynamic VaR combination technique proposed by \cite{bernardi_etal.2014} which average across the models belonging to the SSM. Table \ref{tab:var_comparison} reports three VaR backtesting measures. The first is the Actual over Expected ratio AE, defined as the ratio between the realised VaR exceedances over a given time horizon and their \qmo a priori\qmcsp expected values; VaR forecasts series for which the AE ratio is closer to the unity are preferred. The second and the third backtesting measures are the mean and maximum Absolute Deviation (ADmean and ADmax) of VaR violating returns described in \cite{mcaleer_daveiga.2008}. The ${\rm AD}$ in general provides a measure of the expected loss given a VaR violation; of course models with lower mean and/or maximum ADs are preferred. As showed in Table \ref{tab:var_comparison} the $\mathrm{VaR_{dynamic}}$ series always report lower ADmean and ADmax compared with the $\mathrm{VaR_{avg}}$. Moreover, in two cases (for the SXW1E and the SXXP indices) also the AE ratio is improved.
\begin{table}[!t]
\centering
\begin{tabular}{lccccc}
\toprule
  & \multicolumn{4}{c}{Asset}\\
 & SXA1E & SXP1E & SXW1E & SXXP \\
 \hline
 & \multicolumn{4}{c}{Models} \\
\hline
eGARCH & 7 & 8 & 7 & 5 \\
sGARCH & 16 & 16 & 0 & 0 \\
gjrGARCH & 16 & 16 & 16 & 0 \\
iGARCH & 12 & 9 & 0 & 0 \\
apARCH & 16 & 16 & 16 & 10 \\
csGARCH & 12 & 16 & 0 & 0 \\
TGARCH & 13 & 13 & 14 & 14 \\
AVGARCH & 15 & 15 & 7 & 11 \\
NGARCH & 15 & 16 & 0 & 0 \\
NAGARCH & 16 & 16 & 13 & 13 \\
\hline
 & \multicolumn{4}{c}{Distributions} \\
\hline
norm & 12 & 19 & 5 & 0 \\
snorm & 16 & 18 & 9 & 8 \\
std & 16 & 18 & 10 & 5 \\
sstd & 19 & 17 & 11 & 9 \\
ged & 18 & 18 & 8 & 5 \\
sged & 19 & 18 & 11 & 9 \\
ghyp & 19 & 16 & 9 & 9 \\
jsu & 19 & 17 & 10 & 8 \\
\hline
  & \multicolumn{4}{c}{In--Mean Specification} \\
\hline
Not in mean & 72 & 76 & 43 & 29 \\
in mean & 66 & 65 & 30 & 24 \\
\hline
Total Number & 138 & 141 & 73 & 53 \\
\bottomrule
\end{tabular}
\caption{\footnotesize{Composition of remaining models in the Superior Set for each index.}}
\label{tab:Composition of remaining models}
\end{table}

\begin{table}[!t]
\centering
\begin{tabular}{lccccccc}
\toprule
  & \multicolumn{3}{c}{$\mathrm{VaR_{Dyn}}$} & \multicolumn{3}{c}{$\mathrm{VaR_{Avg}}$}\\
  \cmidrule(lr){1-1} \cmidrule(lr){2-4}\cmidrule(lr){5-7}
 Asset & AE & ADmean & ADmax & AE & ADmean & ADmax \\
 \cmidrule(lr){1-1}\cmidrule(lr){2-4}\cmidrule(lr){5-7}
SXA1E &  1.35  &  0.633  &  2.658  &  1.35  &  0.653  &  2.693 \\
SXP1E &  1.00  &  0.973  &  3.630  &  1.00  &  0.975  &  3.638 \\
SXW1E &  1.20  &  0.406  &  1.936  &  1.20  &  0.466  &  1.954 \\
SXXP &  1.25  &  0.533  &  2.201  &  1.35  &  0.568  &  2.616 \\
\bottomrule
\end{tabular}
\caption{\footnotesize{VaR backtesting measures of the dynamic VaR combination $\rm VaR_{Dyn}$ and the static average $\rm VaR_{Avg}$.}}
\label{tab:var_comparison}
\end{table}
\section{Conclusion}
\label{sec:conclusion}
%:::::::::::::::::::::::::::::::::::::::::::::::::::::::::::::::::::::
%
In the previous Sections we have illustrated the main features of the \proglang{R} package \code{MCS} which implements the Model Confidence Set procedure introduced by \cite{hansen_etal.2011}. The technique proposed by \cite{hansen_etal.2011} is especially useful when more models are available and it is not obvious which one is the best. The MCS sequence of tests delivers the Superior Set of Models (SSM) having Equal Predictive Ability (EPA) in terms of an user supplied loss function discriminating models with respect to desired model characteristics such as, for example, forecasts performances.
%The delivered SSM, $\mathrm{\hat{M}}_{1-\alpha}^{*}$, have the same forecasting ability under the chosen loss function
The \code{MCS} package is very flexible in the types of model and loss function that can be specified by the researcher. This freedom allows the user to concentrate on substantive issues, such as the construction of the initial set of model's specifications $\mathrm{M}^{0}$, without being limited by the constraints imposed by the software. An empirical example shows the relevance of the package by illustrating in details the use of the provided functions. In particular, the example compares the ability of different models belonging to the ARCH family to predict large financial losses and discuss the ARCH--type models and their maximum likelihood estimation using the popular \proglang{R} package \code{rugarch} developed by \cite{ghalanos.2014}. The Model Confidence Set procedure is firstly performed in order to reduce the number of models, and then we show that accounting for the VaR dynamic model averaging technique of \cite{bernardi_etal.2014} improves the VaR forecast performance.
%
%:::::::::::::::::::::::::::::::::::::::::::::::::::::::::::::::::::::
% SUBSECTION: ACKNOWLEDGMENTS
%:::::::::::::::::::::::::::::::::::::::::::::::::::::::::::::::::::::
\section*{Acknowledgments}
%:::::::::::::::::::::::::::::::::::::::::::::::::::::::::::::::::::::
%
This research is supported by the Italian Ministry of Research PRIN 2013--2015, ``Multivariate Statistical Methods for Risk Assessment'' (MISURA), and by the ``Carlo Giannini Research Fellowship'', the ``Centro Interuniversitario di Econometria'' (CIdE) and ``UniCredit Foundation''. In the development of package \code{MCS} we have benefited from the suggestions and help of several users. In particular, we would like to thank Lea Petrella and Riccardo Sucapane for their constructive comments on previous drafts of this work. Our sincere thanks go to all the developers of \proglang{R} since without their continued effort and support no contributed package would exist.
%%%
%
%:::::::::::::::::::::::::::::::::::::::::::::::::::::::::::::::::::::::
% REFERENCES
%:::::::::::::::::::::::::::::::::::::::::::::::::::::::::::::::::::::::

%:::::::::::::::::::::::::::::::::::::::::::::::::::::::::::::::::::::::
%\bibliographystyle{jss}
\bibliography{references}
%:::::::::::::::::::::::::::::::::::::::::::::::::::::::::::::::::::::::

\end{document}

%% file: mydef.tex
\def\0{\mbox{\bf{0}}}

\def\btheta{\mathbf{\theta}}

\def\sM{\mathsf{M}}

\definecolor{darkross}{rgb}{0.008,0.412,0.471}
\definecolor{middleross}{rgb}{0.012,0.580,0.663}
\definecolor{lightross}{rgb}{0.016,0.749,0.855}
\definecolor{darkblue}{rgb}{0.067,0.008,0.471}
\definecolor{middleblue}{rgb}{0.094,0.012,0.663}
\definecolor{lightblue}{rgb}{0.122,0.016,0.855}
\definecolor{darkpurple}{rgb}{0.471,0.008,0.412}
\definecolor{middlepurple}{rgb}{0.663,0.012,0.580}
\definecolor{lightpurple}{rgb}{0.855,0.016,0.749}
\definecolor{darkbrown}{rgb}{0.471,0.067,0.008}
\definecolor{middlebrown}{rgb}{0.663,0.094,0.012}
\definecolor{lightbrown}{rgb}{0.855,0.122,0.016}
\definecolor{darkolive}{rgb}{0.412,0.471,0.008}
\definecolor{middleolive}{rgb}{0.580,0.663,0.012}
\definecolor{lightolive}{rgb}{0.749,0.855,0.016}
\definecolor{darkgreen}{rgb}{0.008,0.417,0.067}
\definecolor{middlegreen}{rgb}{0.012,0.663,0.094}
\definecolor{lightgreen}{rgb}{0.016,0.855,0.122}
\definecolor{darkocre}{rgb}{0.471,0.298,0.008}
\definecolor{middleocre}{rgb}{0.663,0.420,0.012}
\definecolor{lightocre}{rgb}{0.855,0.541,0.016}

\def\qmo{``}
\def\qmc{''}
\def\qmcsp{'' }

%% file: paper_MCS_JSS_v3_arxiv.bbl
\begin{thebibliography}{50}
\newcommand{\enquote}[1]{``#1''}
\providecommand{\natexlab}[1]{#1}
\providecommand{\url}[1]{\texttt{#1}}
\providecommand{\urlprefix}{URL }
\expandafter\ifx\csname urlstyle\endcsname\relax
  \providecommand{\doi}[1]{doi:\discretionary{}{}{}#1}\else
  \providecommand{\doi}{doi:\discretionary{}{}{}\begingroup
  \urlstyle{rm}\Url}\fi
\providecommand{\eprint}[2][]{\url{#2}}

\bibitem[{Barndorff-Nielsen(1977)}]{barndorff.1977}
Barndorff-Nielsen O (1977).
\newblock \enquote{Exponentially decreasing distributions for the logarithm of
  particle size.}
\newblock \emph{Proceedings of the Royal Society of London. A. Mathematical and
  Physical Sciences}, \textbf{353}(1674), 401--419.

\bibitem[{Bates and Granger(1969)}]{bates_granger.1969}
Bates JM, Granger CW (1969).
\newblock \enquote{The combination of forecasts.}
\newblock \emph{OR}, pp. 451--468.

\bibitem[{Bauwens \emph{et~al.}(2006)Bauwens, Laurent, and
  Rombouts}]{bauwens_etal.2006}
Bauwens L, Laurent S, Rombouts JVK (2006).
\newblock \enquote{Multivariate GARCH models: a survey.}
\newblock \emph{Journal of Applied Econometrics}, \textbf{21}(1), 79--109.
\newblock ISSN 1099-1255.
\newblock \doi{10.1002/jae.842}.
\newblock \urlprefix\url{http://dx.doi.org/10.1002/jae.842}.

\bibitem[{{Bernardi} \emph{et~al.}(2014){Bernardi}, {Catania}, and
  {Petrella}}]{bernardi_etal.2014}
{Bernardi} M, {Catania} L, {Petrella} L (2014).
\newblock \enquote{{Are News Important to Predict Large Losses ?}}
\newblock \emph{Working Paper, Arxiv Preprint}.
\newblock \eprint{1306.2834}.

\bibitem[{Black(1976)}]{black.1976}
Black F (1976).
\newblock \enquote{Studies of Stock Price Volatility Changes.}
\newblock \emph{In Proceedings of the 1976 American Statistical Association,
  Business and Economical Statistics Section, Alexandria, VA: American
  Statistical Association}, pp. 177--181.

\bibitem[{Bollerslev(1986)}]{bollerslev.1986}
Bollerslev T (1986).
\newblock \enquote{Generalized autoregressive conditional heteroskedasticity.}
\newblock \emph{Journal of Econometrics}, \textbf{31}(3), 307 -- 327.
\newblock ISSN 0304-4076.
\newblock \doi{http://dx.doi.org/10.1016/0304-4076(86)90063-1}.
\newblock
  \urlprefix\url{http://www.sciencedirect.com/science/article/pii/0304407686900631}.

\bibitem[{Bollerslev(2008)}]{bollerslev.2008}
Bollerslev T (2008).
\newblock \enquote{{Glossary to ARCH (GARCH)}.}
\newblock \emph{CREATES Research Papers 2008-49}, School of Economics and
  Management, University of Aarhus.
\newblock \urlprefix\url{http://ideas.repec.org/p/aah/create/2008-49.html}.

\bibitem[{Bollerslev \emph{et~al.}(1994)Bollerslev, Engle, and
  Nelson}]{bollerslev_etal.1994}
Bollerslev T, Engle RF, Nelson DB (1994).
\newblock \enquote{ARCH models.}
\newblock \emph{Handbook of econometrics}, \textbf{4}, 2959--3038.

\bibitem[{Box and Cox(1964)}]{box_cox.1964}
Box GE, Cox DR (1964).
\newblock \enquote{An analysis of transformations.}
\newblock \emph{Journal of the Royal Statistical Society. Series B
  (Methodological)}, pp. 211--252.

\bibitem[{Clark and McCracken(2001)}]{clark_mccracken.2001}
Clark TE, McCracken MW (2001).
\newblock \enquote{Tests of equal forecast accuracy and encompassing for nested
  models.}
\newblock \emph{Journal of econometrics}, \textbf{105}(1), 85--110.

\bibitem[{Creal \emph{et~al.}(2013)Creal, Koopman, and Lucas}]{creal_etal.2013}
Creal D, Koopman SJ, Lucas A (2013).
\newblock \enquote{Generalized autoregressive score models with applications.}
\newblock \emph{Journal of Applied Econometrics}, \textbf{28}(5), 777--795.

\bibitem[{Diebold and Lopez(1996)}]{diebold_lopez.1996}
Diebold FX, Lopez JA (1996).
\newblock \enquote{Forecast evaluation and combination.}

\bibitem[{Diebold and Mariano(2002)}]{diebold_mariano.2002}
Diebold FX, Mariano RS (2002).
\newblock \enquote{Comparing predictive accuracy.}
\newblock \emph{Journal of Business \& economic statistics}, \textbf{20}(1).

\bibitem[{Ding \emph{et~al.}(1993)Ding, Granger, and Engle}]{ding_etal.1993}
Ding Z, Granger CW, Engle RF (1993).
\newblock \enquote{A long memory property of stock market returns and a new
  model.}
\newblock \emph{Journal of Empirical Finance}, \textbf{1}(1), 83 -- 106.
\newblock ISSN 0927-5398.
\newblock \doi{http://dx.doi.org/10.1016/0927-5398(93)90006-D}.
\newblock
  \urlprefix\url{http://www.sciencedirect.com/science/article/pii/092753989390006D}.

\bibitem[{Engle(1982)}]{engle.1982}
Engle RF (1982).
\newblock \enquote{Autoregressive Conditional Heteroscedasticity with Estimates
  of the Variance of United Kingdom Inflation.}
\newblock \emph{Econometrica}, \textbf{50}(4), 987--1007.
\newblock \urlprefix\url{http://www.jstor.org/stable/1912773}.

\bibitem[{Engle and Bollerslev(1986)}]{engle_bollerslev.1986}
Engle RF, Bollerslev T (1986).
\newblock \enquote{Modelling the persistence of conditional variances.}
\newblock \emph{Econometric Reviews}, \textbf{5}(1), 1--50.
\newblock \doi{10.1080/07474938608800095}.
\newblock \eprint{http://dx.doi.org/10.1080/07474938608800095},
  \urlprefix\url{http://dx.doi.org/10.1080/07474938608800095}.

\bibitem[{Engle and Lee(1993)}]{engle_lee.1993}
Engle RF, Lee GG (1993).
\newblock \enquote{A permanent and transitory component model of stock return
  volatility.}
\newblock \emph{University of California at San Diego, Economics Working Paper
  Series}.

\bibitem[{Engle \emph{et~al.}(1987)Engle, Lilien, and Robins}]{engle_etal.1987}
Engle RF, Lilien DM, Robins RP (1987).
\newblock \enquote{Estimating Time Varying Risk Premia in the Term Structure:
  The Arch-M Model.}
\newblock \emph{Econometrica}, \textbf{55}(2), 391 -- 407.
\newblock \urlprefix\url{http://www.jstor.org/stable/1913242}.

\bibitem[{Fern{\'a}ndez and Steel(1998)}]{fernandez_steel.1998}
Fern{\'a}ndez C, Steel MF (1998).
\newblock \enquote{On Bayesian modeling of fat tails and skewness.}
\newblock \emph{Journal of the American Statistical Association},
  \textbf{93}(441), 359--371.

\bibitem[{Francq and Zakoian(2011)}]{francq_zakoian.2011}
Francq C, Zakoian JM (2011).
\newblock \emph{GARCH models: structure, statistical inference and financial
  applications}.
\newblock John Wiley \& Sons.

\bibitem[{Gallant \emph{et~al.}(1997)Gallant, Hsieh, and
  Tauchen}]{gallant_etal.1997}
Gallant A, Hsieh D, Tauchen G (1997).
\newblock \enquote{Estimation of stochastic volatility models with
  diagnostics.}
\newblock \emph{Journal of Econometrics}, \textbf{81}(1), 159 -- 192.
\newblock ISSN 0304-4076.
\newblock \doi{http://dx.doi.org/10.1016/S0304-4076(97)00039-0}.
\newblock
  \urlprefix\url{http://www.sciencedirect.com/science/article/pii/S0304407697000390}.

\bibitem[{Ghalanos(2014)}]{ghalanos.2014}
Ghalanos A (2014).
\newblock \emph{rugarch: Univariate GARCH models.}
\newblock R package version 1.3-1.

\bibitem[{Giacomini and White(2006)}]{giacomini_white.2006}
Giacomini R, White H (2006).
\newblock \enquote{Tests of conditional predictive ability.}
\newblock \emph{Econometrica}, \textbf{74}(6), 1545--1578.

\bibitem[{Glosten \emph{et~al.}(1993)Glosten, Jagannathan, and
  Runkle}]{glosten_etal.1993}
Glosten LR, Jagannathan R, Runkle DE (1993).
\newblock \enquote{On the Relation between the Expected Value and the
  Volatility of the Nominal Excess Return on Stocks.}
\newblock \emph{The Journal of Finance}, \textbf{48}(5), 1779--1801.
\newblock ISSN 1540-6261.
\newblock \doi{10.1111/j.1540-6261.1993.tb05128.x}.
\newblock \urlprefix\url{http://dx.doi.org/10.1111/j.1540-6261.1993.tb05128.x}.

\bibitem[{Gonz\'{a}lez-Rivera \emph{et~al.}(2004)Gonz\'{a}lez-Rivera, Lee, and
  Mishra}]{gonzalez_etal.2004}
Gonz\'{a}lez-Rivera G, Lee TH, Mishra S (2004).
\newblock \enquote{Forecasting volatility: A reality check based on option
  pricing, utility function, value-at-risk, and predictive likelihood.}
\newblock \emph{International Journal of Forecasting}, \textbf{20}(4), 629 --
  645.
\newblock ISSN 0169-2070.
\newblock \doi{http://dx.doi.org/10.1016/j.ijforecast.2003.10.003}.
\newblock
  \urlprefix\url{http://www.sciencedirect.com/science/article/pii/S0169207003001420}.

\bibitem[{Hansen(2005)}]{hansen_reinhard.2005}
Hansen PR (2005).
\newblock \enquote{A test for superior predictive ability.}
\newblock \emph{Journal of Business \& Economic Statistics}, \textbf{23}(4).

\bibitem[{Hansen and Lunde(2005)}]{hansen_lunde.2005}
Hansen PR, Lunde A (2005).
\newblock \enquote{A forecast comparison of volatility models: does anything
  beat a GARCH(1,1)?}
\newblock \emph{Journal of Applied Econometrics}, \textbf{20}(7), 873--889.
\newblock ISSN 1099-1255.
\newblock \doi{10.1002/jae.800}.
\newblock \urlprefix\url{http://dx.doi.org/10.1002/jae.800}.

\bibitem[{Hansen \emph{et~al.}(2003)Hansen, Lunde, and
  Nason}]{hansen_etal.2003}
Hansen PR, Lunde A, Nason JM (2003).
\newblock \enquote{Choosing the best volatility models: The model confidence
  set approach.}
\newblock \emph{Oxford Bulletin of Economics and Statistics}, \textbf{65}(s1),
  839--861.

\bibitem[{Hansen \emph{et~al.}(2011)Hansen, Lunde, and
  Nason}]{hansen_etal.2011}
Hansen PR, Lunde A, Nason JM (2011).
\newblock \enquote{The model confidence set.}
\newblock \emph{Econometrica}, \textbf{79}(2), 453--497.

\bibitem[{Harvey(2013)}]{harvey.2013}
Harvey AC (2013).
\newblock \emph{Dynamic Models for Volatility and Heavy Tails: With
  Applications to Financial and Economic Time Series}.
\newblock 52. Cambridge University Press.

\bibitem[{Harvey and Shephard(1996)}]{harvey_shephard.1996}
Harvey AC, Shephard N (1996).
\newblock \enquote{Estimation of an Asymmetric Stochastic Volatility Model for
  Asset Returns.}
\newblock \emph{Journal of Business \& Economic Statistics}, \textbf{14}(4),
  429--434.
\newblock \doi{10.1080/07350015.1996.10524672}.
\newblock
  \eprint{http://amstat.tandfonline.com/doi/pdf/10.1080/07350015.1996.10524672},
  \urlprefix\url{http://amstat.tandfonline.com/doi/abs/10.1080/07350015.1996.10524672}.

\bibitem[{Higgins and Bera(1992)}]{higgins_bera.1992}
Higgins ML, Bera AK (1992).
\newblock \enquote{A class of nonlinear ARCH models.}
\newblock \emph{International Economic Review}, pp. 137--158.

\bibitem[{Kilian(1999)}]{kilian.1999}
Kilian L (1999).
\newblock \enquote{Exchange rates and monetary fundamentals: What do we learn
  from long-horizon regressions?}
\newblock \emph{Journal of Applied Econometrics}, \textbf{14}(5), 491--510.
\newblock \urlprefix\url{http://hdl.handle.net/2027.42/34956}.

\bibitem[{Lopez(2001)}]{lopez.2001}
Lopez JA (2001).
\newblock \enquote{Evaluating the predictive accuracy of volatility models.}
\newblock \emph{Journal of Forecasting}, \textbf{20}(2), 87--109.

\bibitem[{McAleer and da~Veiga(2008)}]{mcaleer_daveiga.2008}
McAleer M, da~Veiga B (2008).
\newblock \enquote{Single-index and portfolio models for forecasting
  value-at-risk thresholds.}
\newblock \emph{Journal of Forecasting}, \textbf{27}(3), 217--235.
\newblock ISSN 1099-131X.
\newblock \doi{10.1002/for.1054}.
\newblock \urlprefix\url{http://dx.doi.org/10.1002/for.1054}.

\bibitem[{Nelson(1991)}]{nelson.1991}
Nelson DB (1991).
\newblock \enquote{{Conditional Heteroskedasticity in Asset Returns: A New
  Approach}.}
\newblock \emph{Econometrica}, \textbf{59}(2), 347--70.
\newblock
  \urlprefix\url{http://ideas.repec.org/a/ecm/emetrp/v59y1991i2p347-70.html}.

\bibitem[{{R Development Core Team}(2013)}]{R.2013}
{R Development Core Team} (2013).
\newblock \emph{R: A Language and Environment for Statistical Computing}.
\newblock R Foundation for Statistical Computing, Vienna, Austria.
\newblock {ISBN} 3-900051-07-0, \urlprefix\url{http://www.R-project.org/}.

\bibitem[{Rigby and Stasinopoulos(2005)}]{rigby_stasinopoulos.2005}
Rigby R, Stasinopoulos D (2005).
\newblock \enquote{Generalized additive models for location, scale and shape.}
\newblock \emph{Journal of the Royal Statistical Society: Series C (Applied
  Statistics)}, \textbf{54}(3), 507--554.

\bibitem[{Romano and Wolf(2005)}]{romano_wolf.2005}
Romano JP, Wolf M (2005).
\newblock \enquote{Stepwise multiple testing as formalized data snooping.}
\newblock \emph{Econometrica}, \textbf{73}(4), 1237--1282.

\bibitem[{Samuels and Sekkel(2011)}]{samuels_sekkel.2011}
Samuels JD, Sekkel RM (2011).
\newblock \enquote{Forecasting with Large Datasets: Trimming Predictors and
  Forecast Combination.}
\newblock \emph{Working paper}.

\bibitem[{Samuels and Sekkel(2013)}]{samuels_etal.2013}
Samuels JD, Sekkel RM (2013).
\newblock \enquote{Forecasting with many models: Model confidence sets and
  forecast combination.}
\newblock \emph{Technical report}, Bank of Canada Working Paper.

\bibitem[{Schwert(1990)}]{schwert.1990}
Schwert G (1990).
\newblock \enquote{Stock volatility and the crash of '87.}
\newblock \emph{Review of Financial Studies}, \textbf{3}(1), 77--102.
\newblock \doi{10.1093/rfs/3.1.77}.
\newblock \eprint{http://rfs.oxfordjournals.org/content/3/1/77.full.pdf+html},
  \urlprefix\url{http://rfs.oxfordjournals.org/content/3/1/77.abstract}.

\bibitem[{Silvennoinen and Ter{\"a}svirta(2009)}]{silvennoinen_terasvirta.2009}
Silvennoinen A, Ter{\"a}svirta T (2009).
\newblock \enquote{Multivariate GARCH models.}
\newblock In \emph{Handbook of Financial Time Series}, pp. 201--229. Springer.

\bibitem[{Taylor(1986)}]{taylor.1986}
Taylor SJ (1986).
\newblock \emph{Modelling Financial Times Series}.
\newblock Wiley.

\bibitem[{Taylor(1994)}]{taylor.1994}
Taylor SJ (1994).
\newblock \enquote{MODELING STOCHASTIC VOLATILITY: A REVIEW AND COMPARATIVE
  STUDY.}
\newblock \emph{Mathematical Finance}, \textbf{4}(2), 183--204.
\newblock ISSN 1467-9965.
\newblock \doi{10.1111/j.1467-9965.1994.tb00057.x}.
\newblock \urlprefix\url{http://dx.doi.org/10.1111/j.1467-9965.1994.tb00057.x}.

\bibitem[{Ter{\"a}svirta(2009)}]{terasvirta.2009}
Ter{\"a}svirta T (2009).
\newblock \enquote{An introduction to univariate GARCH models.}
\newblock In \emph{Handbook of Financial Time Series}, pp. 17--42. Springer.

\bibitem[{West(1996)}]{west.1996}
West KD (1996).
\newblock \enquote{Asymptotic inference about predictive ability.}
\newblock \emph{Econometrica: Journal of the Econometric Society}, pp.
  1067--1084.

\bibitem[{White(2000)}]{white.2000}
White H (2000).
\newblock \enquote{A reality check for data snooping.}
\newblock \emph{Econometrica}, \textbf{68}(5), 1097--1126.

\bibitem[{Wuertz \emph{et~al.}(2013)Wuertz, with contribution~from
  Michal~Miklovic, Boudt, Chausse, and {others}}]{fgarch}
Wuertz D, with contribution~from Michal~Miklovic YC, Boudt C, Chausse P,
  {others} (2013).
\newblock \emph{fGarch: Rmetrics - Autoregressive Conditional Heteroskedastic
  Modelling}.
\newblock R package version 3010.82,
  \urlprefix\url{http://CRAN.R-project.org/package=fGarch}.

\bibitem[{Zakoian(1994)}]{zakoian.1994}
Zakoian JM (1994).
\newblock \enquote{Threshold heteroskedastic models.}
\newblock \emph{Journal of Economic Dynamics and Control}, \textbf{18}(5), 931
  -- 955.
\newblock ISSN 0165-1889.
\newblock \doi{http://dx.doi.org/10.1016/0165-1889(94)90039-6}.
\newblock
  \urlprefix\url{http://www.sciencedirect.com/science/article/pii/0165188994900396}.

\end{thebibliography}
